\documentclass[aps,groupedaddress,showpacs,epsfig,floats]{revtex4}
\usepackage{graphicx}
\begin{document}
\title{Magnetically induced explosion of a giant vortex state in a mesoscopic superconducting disk}
\author{D. S. Golubovi\'{c}$^{\,*}$, M. V. Milo\v{s}evi\'{c}$^{\,\S}$, F. M. Peeters$^{\,\S}$ and V. V. Moshchalkov$^{\,*}$ }
\affiliation{$^{*\,}$Nanoscale Superconductivity and Magnetism
Group, Laboratory for Solid State Physics and Magnetism, K. U.
Leuven, Celestijnenlaan 200 D, B-3001 Leuven,
Belgium\\$^{\S\,}$Departement Fysica, Universiteit Antwerpen
(Campus Drie Eiken), Universiteitsplein 1, B-2610 Antwerpen,
Belgium }

\begin{abstract}
The nucleation of superconductivity in a superconducting disk with
a Co/Pt magnetic triangle was studied. We demonstrate that when
the applied magnetic field is parallel to the magnetization of the
triangle, the giant vortex state of vorticity three splits into
three individual $\Phi _{0}$-vortices, due to a pronounced
influence of the C$_{3}$ symmetry of the magnetic triangle. As a
result of a strong pinning of the three vortices by the triangle,
their configuration remains stable in a broad range of applied
magnetic fields. For sufficiently high fields, $\Phi
_{0}$-vortices merge and the nucleation occurs through the giant
vortex state. The theoretical analysis of this novel reentrant
behaviour at the phase boundary, obtained within the Ginzburg -
Landau formalism, is in excellent agreement with the experimental
data.
\end{abstract} 
\maketitle

The nucleation of superconductivity in mesoscopic samples, whose
dimensions are comparable to the superconducting coherence length
$\xi(T)$ and the penetration depth $\lambda (T)$, is substantially
affected by the sample boundary (see Ref. \cite{vvm} and
references therein). It is well established that in circular
mesoscopic disks and loops the onset of superconductivity mostly
occurs through the giant vortex state (GVS) due to their
cylindrical symmetry \cite{d1}. On the other hand, in
superconducting squares and triangles for certain magnetic fields
the GVS easily splits into individual $\Phi _{0}$-vortices ($\Phi
_{0}$ is the superconducting flux quantum), with a possible
generation of additional antivortices \cite{t1}. The transition
from the GVS to a set of $\Phi _{0}$-vortices is caused by the
reduced axial symmetry of squares and triangles and ensures that a
vortex pattern conforms to the symmetry imposed by the boundary of
the sample.

In this paper we investigate the onset of superconductivity in a
mesoscopic  disk on which a magnetic triangle with out-of-plane
magnetization is placed. The nucleation of superconductivity in
hybrid superconductor/ferromagnet disks and loops has been studied
previously, both experimentally and theoretically, as these are a
good model system to gain insight into the interplay between
superconductivity and magnetism in the context of magnetic vortex
pinning and related phenomena
\cite{f1,f2,f3,magpin}. 
In this paper we show that the {\it symmetry} of the magnetic
triangle has a profound effect on the onset of superconductivity
in the disk. Due to the competition between the cylindrical
symmetry of the superconducting disk and the $C_{3}$ symmetry of
the stray field, superconductivity nucleates as a GVS for
vorticity $L=2$, whereas for $L=3$ the GVS splits into three $\Phi
_{0}$-vortices. As a result of a strong pinning of the three
individual vortices their configuration remains stable for $L=4$
and $L=5$.
\begin{figure}[htb] \centering
\vspace{-0.8cm}
\includegraphics*[width=8.9cm]{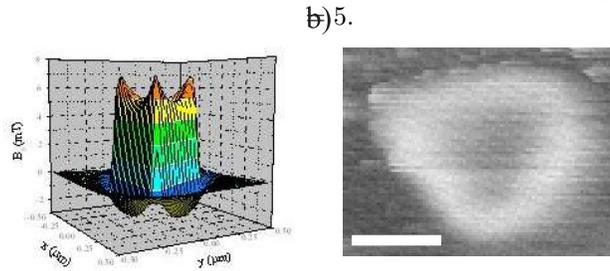}
\vspace{-0.5cm} \caption{The calculated spatial profile (a) of the
stray field and (b) a magnetic force micrograph of the triangle.
The field values were calculated using the saturation
magnetization of bulk Co.\label{stray}}
\end{figure}

The sample was prepared by electron beam lithography, using double
resist technique and lift-off in two steps. For details we refer
to Ref. \cite{f2}. In the first step $30\,$nm Al disks were
prepared using thermal evaporation. Upon the alignment, in the
second step Co/Pt magnetic triangles with the designed size of
$650\,$nm were grown by electron beam evaporation on the disks.
Some disks were left without magnetic elements to be used as
reference samples. The Co/Pt magnetic triangle consists of a
$2.5\,$nm Pt buffer layer a 10 bi-layers of Co and Pt with the
thicknesses of $0.4\,$nm and $1\,$nm, respectively. The triangle
is separated from the superconducting disk by a $2\,$nm Si spacer
layer to avoid the proximity effect and suppression of the order
parameter in the disk below the magnetic triangle. Co/Pt
multi-layers are known to have a pronounced perpendicular
(out-of-plane) anisotropy, caused by the surface anisotropy
between Co and Pt layers, and a high remanence \cite{mag}. The
coercive field of the co-evaporated reference film is $170\,$mT at
room temperature, with a $90\,\%$ remanence. The calculated
spatial profile of the stray field $B_{z}(x,y)$ generated by the
triangle is shown in Fig. \ref{stray}. Prior to the measurements
the triangle was saturated along the easy axis in a high magnetic
field. Therefore, it has been assumed that
 applied magnetic fields in the range $\pm 15\,$mT did
not affect the magnetization of the triangle during the
measurements.

\begin{figure}[htb] \centering
\includegraphics*[width=6.5cm]{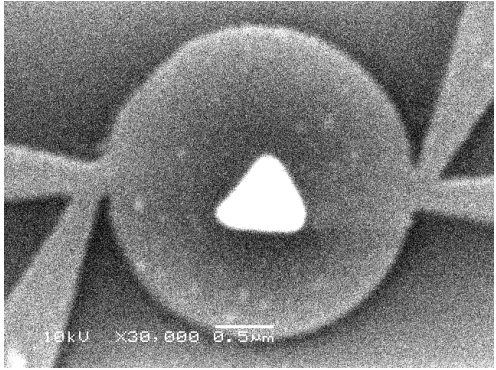}
\caption{A scanning electron micrograph of the
structure.\label{sample}}
\end{figure}

A scanning electron micrograph of the structure is shown in Fig.
\ref{sample}. The radius of the disk is $1.34\,{\rm \mu m}$, which
exceeds approximately $3\,\%$ the patterned radius. Even though
the triangle was designed as equilateral, its sides are $690\,$nm,
$650\,$nm and $740\,$nm, due to a relatively low acceleration
voltage and a lack of the proximity correction tool in our e-beam
system.

The superconducting $T_{c}(B)$ phase boundary was obtained by
four-point transport measurements in a $^{4}$He cryogenic setup at
temperatures down to $1.25\,$K with the temperature stability of
$0.4\,$mK, applying the magnetic field perpendicularly to the
sample surface. A transport current with the effective value of
$100\,$nA and frequency $27.7\,$Hz was used. The measurements were
taken by sweeping the magnetic field at a constant temperature.
The temperature and field steps were $0.5\,$mK and $50\,{\rm \mu
T}$, respectively. A special attention was being paid to eliminate
any possible trapped flux in the set-up during the measurements.
The room temperature resistance and the residual resistance at
$4.2\,$ K of the structure shown in Fig. \ref{sample} are
$6.5\,{\rm \Omega}$ and $3.2\,{\rm \Omega}$, respectively. The
mean free path of Al, estimated from the co-evaporated reference
film, is $12.3\,$nm, so that the samples are in the dirty limit
with the coherence length $\xi (0) = 120\,$nm. The critical
temperature of the disk with the triangle in zero applied field is
$T_{c0}=1.4136\,$K, whereas for the reference disk is
$T_{c0}^{(R)}=1.4366\,$K.

Hereafter we will be referring to a magnetic field applied
parallel to the magnetization of the triangle as positive.

\begin{figure}[htb] \centering
\vspace{-0.8cm}
\includegraphics*[width=9cm]{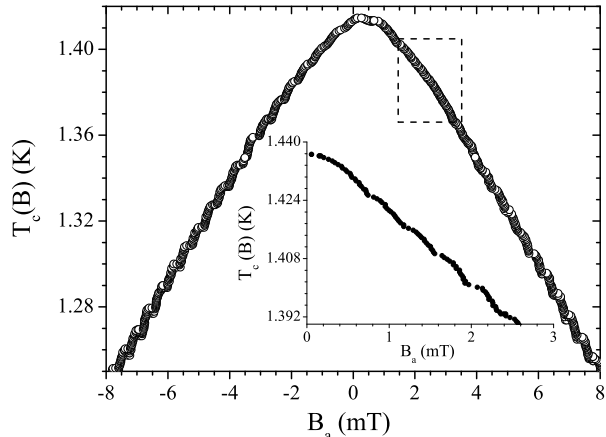}
\vspace{-1.cm} \caption{Experimental $T_{c}(B)$ phase boundary of
a superconducting disk with a  magnetic triangle  on top. Dashed
square indicates the temperature region where no cusps in the
phase boundary are observed. The inset shows a part of the
$T_{c}(B)$ phase boundary of a reference disk down to $1.39\,$K.
\label{tcbexp}}
\end{figure}

The experimental superconducting $T_{c}(B)$ phase boundary is
shown is Fig. \ref{tcbexp}. The inset is a part of the $T_{c}(B)$
phase boundary of a reference superconducting disk, without the
magnetic triangle, down to $0.967\,T_{c0}^{(R)}$. The shift of the
$T_{c}(B)$ phase boundary along the field axis, with the maximum
critical temperature $T_{cm}$ attained for a finite value of an
applied magnetic field, was already previously observed and is
caused by the compensation of the stray field generated by the dot
\cite{f2,f3}.
It can be seen that for negative applied fields the $T_{c}(B)$
phase boundary exhibits a typical cusp-like behaviour in the whole
temperature range, with each cusp corresponding to a change in
vorticity by one. However, for positive applied fields there are
no cusps in the $T_{c}(B)$ phase boundary for a broad range of 
temperatures between $0.991\,T_{cm}$ and $0.969\,T_{cm}$. The
region is indicated by a dashed square in Fig. \ref{tcbexp}. For
lower temperatures, the cusp-like behaviour is recovered. In the
same relative temperature range, the $T_{c}(B)$ phase boundary of
a reference superconducting disk exhibits the typical behaviour
\cite{w}.

The experimental data were analyzed in the framework of the
Ginzburg-Landau (GL) theory. The 3D GL equations were solved
numerically, without imposing any constraints on the final form of
the order parameter at a given temperature and magnetic field (for
details of the approach, see Ref. \cite{schw} and references
therein). At the superconductor/vacuum interface the Neumann
boundary condition for the order parameter was used. The influence
of the current and voltage contacts on the nucleation process  was
modelled by adding two stripes with length of $400\,$nm and width
of $250\,$nm to the disk, at the position of the current contacts.
Even though they geometrically differ from the real contacts, they
allow for a slight local spread of the screening currents in the
disk, as well as provide a local enhancement of the
superconductivity in the contact area. Therefore, they may be used
to adequately describe the physical phenomena related to the real
contacts. The simulations were carried out with the exact
dimensions of the disk and triangle, using the saturation
magnetization of bulk Co $m=140\,$mT, with the coherence length as
a fitting parameter. The experimental results were theoretically
best reproduced  with $\xi(0)=118\,$nm, which is in excellent
agreement with the coherence length obtained for the reference
film.

\begin{figure}[htb] \centering
\vspace{-0.5cm}
\includegraphics*[width=9.5cm]{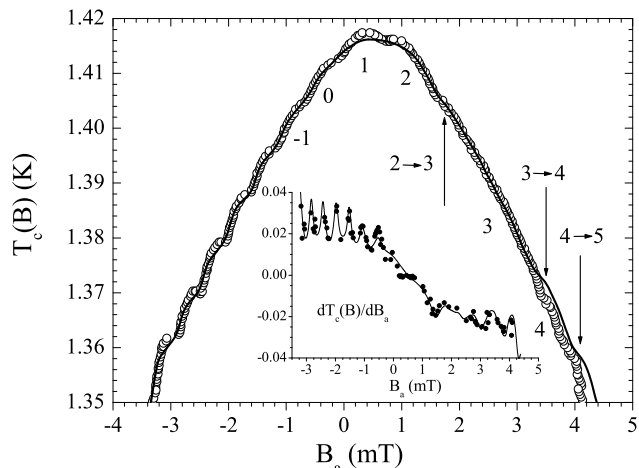}
\vspace{-1cm} \caption{$T_{c}(B)$ phase boundary of a
superconduting disk with a magnetic triangle. The solid line
presents the theoretical curve, whereas open symbols show
experimental data. Numbers stand for the vorticity, with arrows
indicating the transitions between the states with different
vorticities. The inset shows the $dT_{c}(B)/dB_{a}$ versus the
applied field. Filled symbols are experimental data and line is
the theoretical curve. \label{tcb}}
\end{figure}

\begin{figure}[htb] \centering
\vspace{-1.3cm}
\includegraphics*[width=7.5cm]{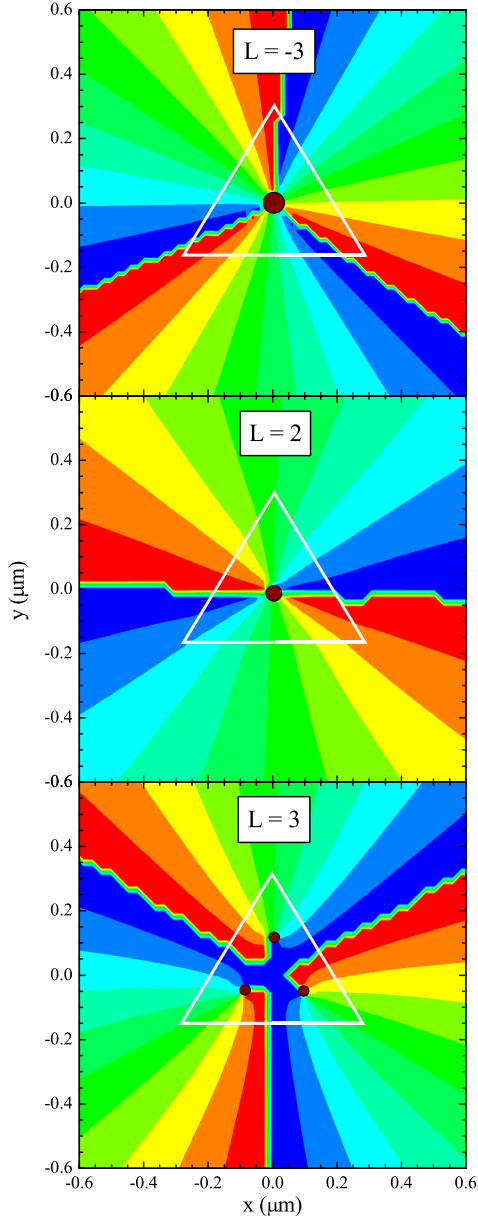}
\vspace{-0.8cm} \caption{The contour plot of the phase of the
order parameter for the vorticities $L=-3$, $L=2$ and $L=3$.
Blue/red colour are $0/2\pi$ of the phase. The white triangle
illustrates the position of the magnetic triangle. \label{phase}}
\end{figure}

Fig. \ref{tcb} shows the experimental and theoretical $T_{c}(B)$
phase boundaries of the disk with the triangle, down to
$0.95\,T_{cm}$. Open symbols are the experimental data, whereas
the solid line presents the theoretically obtained  phase
boundary. The numbers indicate vorticities and arrows point the
field values at which a particular transition occurs.
The inset shows the $dT_{c}(B)/dB_{a}$ versus the applied field
$B_{a}$. The stray field of the triangle creates one vortex in the
disk in the absence of the external magnetic field. The state
without vortices ($L=0$) is, therefore, realized by applying small
negative magnetic fields. For negative applied fields, the
nucleation of superconductivity occurs through the GVS, which is
reflected by a regular cusp-like character of the $T_{c}(B)$ phase
boundary. The contour plot of the phase of the order parameter for
$L=-3$, shown in Fig. \ref{phase}, illustrates the presence of the
GVS.

The GVS is retained in positive applied fields for up to $L=2$.
However, for vorticity $L=3$ the lowest energy state corresponds
to three $\Phi_{0}$-vortices. The contour plots of the phase of
the order parameter for the  $L=2$ and $L=3$ are shown in Fig.
\ref{phase}.
It is clear that the state $L=2$ is a GVS located below the
triangle, whereas for $L=3$ there are three individual vortices,
located towards the apices of the triangle. The three $\Phi _{0}$
vortices are strongly pinned at their positions by the magnetic
triangle. This results in a substantially enhanced stability of
the $L=3$ state, and gives rise to an extended cusp in the
$T_{c}(B)$ phase boundary. The configuration of the three cornered
vortices is imposed by the symmetry and remains stable for $L=4$
and $L=5$. For higher applied fields (and consequently higher
vorticities), vortices are compressed under the triangle at the
centre of the disk and forced by the screening currents to merge
back to the GVS. This reentrant-like behaviour is reflected by the
reappearance of regular cusps in the $T_{c}(B)$ phase boundary for
higher fields (see Fig. \ref{tcbexp}).

Obviously, in addition to the character and symmetry of the
boundary, the onset of superconductivity in hybrid
superconductor/ferromagnet structures is profoundly influenced by
the stray field locally generated by the magnetic element. The
extent of the influence depends on the mutual relation between the
parameters of the superconductor and the ferromagnet. As detailed
in Ref. \cite{misko}, in the theory of vortex pinning by
perpendicularly magnetized dots, the lowest energy state is
achieved when an external vortex is located below the dot (in the
parallel case), whereas an eventual anti-vortex would be repelled
and its equilibrium position is determined by other relevant
parameters. As a general rule, the prevailing mechanism which
governs the vortex behaviour is its interaction with the screening
currents generated by the stray field of a magnetic dot.

For negative applied fields, which generate anti-vortices with
respect to the direction of the magnetization of the triangle, the
symmetry of the triangle does not have a major influence on the
nucleation process. Namely, for such field polarity, the triangle
and external field induce opposite supercurrents in the disk,
which tend to cancel each other. With increasing negative applied
field, the vortex-like currents prevail and after the critical
conditions are reached antivortices nucleate. The onset occurs
through the GVS, as imposed by the cylindrical symmetry of the
disk. For positive applied fields, on the other hand, there exists
a competition between the cylindrical symmetry of the
superconducting disk and the $C_{3}$ symmetry of the magnetic
triangle, which governs the onset. Both currents, induced by the
externally applied field and by the triangle, compress the
vortices to the centre of the disk, but imposing different
geometries on the final vortex configuration. For $L=1$ and $L=2$,
the distribution of the order parameter is influenced by the
symmetry of the magnetic triangle through the suppression of the
order parameter under its edge \cite{misko}, but there is no
particular correspondence between the vortex patterns and the
symmetry of the stray field. For $L=3$, however, the $C_{3}$
symmetry of the triangle has a pronounced influence and the GVS
splits into three $\Phi _{0}$-vortices. This splitting is caused
by the tendency of the system to minimize its kinetic energy,
associated with the screening currents. This process can be
thought of as the competition between the triangularly imposed
confinement, compressing vortices to its centre and their mutual
repulsion. Therefore, due to the $C_{3}$ symmetry, only the state
$L=3$ may actually be stabilized as a collection of $\Phi
_{0}$-vortices. The vortices assume the positions close to the
apices of the triangle, where the positive stray field is the
highest.

Given that the specific distribution of the screening currents
following from the shape of the triangle, the configuration of the
three $\Phi_{0}$ vortices which minimizes the kinetic energy
remains favourable and stable in a broad range of applied fields,
that is, the three vortices are strongly pinned at their
positions. The enhanced temperature stability range follows from
the size constraints of the pinning site. Namely, the fourth
vortex can only enter under the triangle when the pinning area
becomes large enough in terms of $\xi(T)$, so that the pinning
force overwhelms the vortex-repulsion. The fourth vortex is then
pushed to the centre of the triangle, without causing any
substantial rearrangement of the three vortices. For vorticity
$L=5$ three $\Phi _{0}$-vortices remain stable and pinned at the
apices of the triangle, whereas a GVS with vorticity 2 is created
at the centre of the triangle. For higher fields, the vortices
below the triangle merge, forming a single GVS state. This is the
first time that such a reentrant behaviour at the phase boundary
has been observed.

In conclusion, we have investigated the onset of superconductivity
in a mesoscopic superconducting disk with a perpendicularly
magnetized magnetic triangle. We demonstrated that the symmetry of
the magnetic triangle strongly affects the nucleation process. For
applied field oriented antiparallel to the magnetization of the
triangle, the nucleation of superconductivity is predominantly
influenced by the boundary of the disk and occurs through the GVS,
whereas for fields applied parallel to the magnetic moment of the
triangle the GVS with vorticity $L=3$ splits into three $\Phi
_{0}$-vortices. This magnetically induced explosion of the GVS is
generally applicable to mesoscopic superconductors with magnetic
polygons on top, where the number of apices would determine the
particular vorticity at which the GVS splits into a set of
$\Phi_{0}$ vortices.

This work was supported by the Research Fund K. U. Leuven
GOA/2004/02 programme, the University of Antwerp (GOA), the
Flemish FWO and the Belgian IUAP programmes, as well as by the
JSPS/ESF "Nanoscience and Engineering in Superconductivity"
programme.


\begin{thebibliography}{20}
\bibitem{vvm} V. V. Moshchalkov {\it et al.}, Nature {\bf 373},
319 (1995).
\bibitem{d1} V. A. Schweigert, F. M. Peeters and P. S. Deo, Phys.
Rev. Lett. {\bf 81}, 2783 (1998); A. K. Geim {\it et al.}, Nature
(London) {\bf 390}, 259 (1997); V. Bruyndoncx {\it et al.}, Phys.
Rev. B {\bf 60}, 10468 (1999); A. Kanda {\it el al.}, Phys. Rev.
Lett. {\bf 93}, 257002 (2004).
\bibitem{t1} L. F. Chibotaru {\it et al.}, Nature (London) {\bf 408}, 833 (2000);
{\it ibid.} Phys. Rev. Lett. {\bf 86}, 1323 (2001); V. R. Misko
{\it et al.}, Phys. Rev. Lett. {\bf 90}, 147003 (2003).
\bibitem{w} F. M. Peeters and V. A. Schweigert, Phys. Rev. B
{\bf 63}, 144517 (2001); W. V. Pogosov, {\it ibid}. {\bf 65},
224511 (2002).
\bibitem{f1} M. V. Milo\v{s}evi\'{c}, S. V. Yampolskii and F. M.
Peeters, Phys. Rev. B {\bf 66}, 024515 (2002)
\bibitem{f2} D. S. Golubovi\'{c} {\it et al.},
Phys. Rev. B. {\bf 68}, 172503 (2003)
\bibitem{f3} D. S. Golubovi\'{c} {\it et al.},
Europhys. Lett. {\bf 65}, 546 (2004).
\bibitem{magpin} A. Y. Aladyshkin, A. S. Mel'nikov and D. A.
Ryzhov, J. Phys.: Condens. Matter {\bf 38}, 6591 (2003); J. E.
Villegas {\it et al.}, Science {\bf 302}, 1188 (2003); M. Lange
{\it et al.}, Phys. Rev. Lett. {\bf 90}, 197006 (2003).
\bibitem{mag} P. F. Carcia, J. Appl. Phys. {\bf 63}, 10 (1998);
J. V. Harzer {\it et al.}, {\it ibid}. {\bf 69}, 2448 (1991).
\bibitem{schw}  V.A.~Schweigert and F.M.~Peeters, Phys. Rev. B {\bf 57}, 13817 (1998).
\bibitem{misko} M. V. Milo\v{s}evi\'{c} and F. M. Peeters, Phys.
Rev. B {\bf 68}, 094510 (2003); {\it ibid.} Phys. Rev. B {\bf 69},
104522 (2004).
\end{thebibliography}
\end{document}